# Wave-front shaping in nonlinear multimode fibers

OMER TZANG, ANTONIO M. CARAVACA-AGUIRRE, KELVIN WAGNER, RAFAEL PIESTUN

*Department of Electrical, Computer, and Energy Engineering, University of Colorado, Boulder, Colorado 80309, USA*
*omer.tzang@colorado.edu*

**ABSTRACT**

Recent remarkable progress in wave-front shaping has enabled control of light propagation inside linear media to focus and image through scattering objects. In particular, light propagation in multimode fibers comprises complex intermodal interactions and rich spatiotemporal dynamics. Control of physical phenomena in multimode fibers and its applications is in its infancy, opening opportunities to take advantage of complex mode interactions. In this work, we demonstrate a wave-front shaping approach for controlling nonlinear phenomena in multimode fibers. Using a spatial light modulator at the fiber's input and a genetic algorithm optimization, we control a highly nonlinear stimulated Raman scattering cascade and its interplay with four wave mixing via a flexible implicit control on the superposition of modes that are coupled into the fiber. We show for the first time versatile spectrum manipulations including shifts, suppression, and enhancement of Stokes and anti-Stokes peaks. These demonstrations illustrate the power of wave-front shaping to control and optimize nonlinear wave propagation.

## INTRODUCTION

Controlling light propagation through complex media is key in imaging and light energy delivery applications[1]. In the last decade, a renewed interest in the topic was sparked by new fundamental discoveries[2,3] as well as technological improvements in devices such as spatial light modulators (SLM) and computation capabilities. Accordingly, techniques for manipulating the wave-front incident onto the complex medium using high-resolution SLMs have helped mitigate scattering in random media[4,5] and mode dispersion and coupling in multimode optical-fibers (MMF)[6–10]. Recent progress in the understanding of optical nonlinear media[11–13] raises interest in nonlinear propagation for both fundamental and applied research. While nonlinear propagation in single-mode fibers has been thoroughly investigated, only sparse studies have addressed the richer nonlinear pulse propagation in MMFs, leaving this field largely unexplored with opportunities to exploit the multimodal degrees of freedom for controlling multi-dimensional spectral-spatio-temporal interactions[14–17].

Single mode fibers have traditionally been adopted for most nonlinear applications due to the simplicity of their modal structure and propagation dynamics[18]. However, multimode fibers are gaining new interest due to their potential as higher bandwidth waveguides for communication using space-division-multiplexing [19–21] and high-power fiber lasers[22,23]. In fiber lasers, the higher damage threshold of larger fibers is attractive as an alternative for power-limited single-mode fiber lasers and amplifiers[24]. MMF are important for endoscopic nonlinear microscopy and laser surgery, where nonlinear pulse distortions are expected[25]. Recently, control over a variety of spatiotemporal nonlinear dynamics in graded-index (GRIN) MMF has been demonstrated by manually adjusting (laterally shifting) the input beam coupling to the fiber [15,26].

In this work, we introduce wave-front shaping (WFS) to control nonlinear interactions using a SLM at the input coupling of the fiber. Using genetic algorithm (GA) based optimizations, we tailor and optimize the highly nonlinear generation of a stimulated Raman scattering (SRS) cascade and four wave mixing (FWM) in GRIN multimode fibers. Our methodology allows enhancement, suppression and shifting of selected Stokes or anti-Stokes (FWM) peaks by WFS optimization of the mode-superposition at the fiber's input. It should be emphasized that the wave-front feedback control achieved with a SLM cannot be achieved with basic shifts of the laser spatial input coupling or alignment into the fiber. Hence, WFS further provides a systematic approach for controlling and monitoring the complex dynamics of nonlinear phenomena in MMF. Our GA optimization presents a solution to the inverse problem seeking to find the superposition of modes that enhances or suppresses specific nonlinear process. Remarkably, because the process is implemented experimentally online, it inherently takes into account all optical system aberrations, misalignments, and fiber actual configurations.

**Linear wave-front shaping through fibers**

Propagation of light in MMFs comprises a superposition of discrete propagating modes. Phase-velocity mode dispersion is aggravated by random mode coupling arising from imperfections and bends. They all contribute to creating complex 3D interference patterns,[10] which result in a random speckle field at the fiber's output. Image transmission through optical fibers was originally proposed theoretically by Yariv et al[27,28], followed by proof of concept experiments[29,30]. Lately, new ideas utilizing digital phase conjugation[7,31] and WFS[6,8,9,32] have expanded control over the different modes, showing imaging potential for endoscopy applications[7,10,25,33]. Spatial beam optimization in fibers with gain have also been recently studied[34]. Linear WFS techniques are based on an optimization of the input wave-front, an experimental determination of the optical transmission matrix, or on direct phase conjugation. Linearity of the system is a basic assumption in the transmission matrix formalism[35]. In contrast to previous work, here WFS is used in the nonlinear regime where the transmission matrix formalism is not directly applicable. The nonlinear propagation is complex and cannot be described as a linear super-position of uncoupled modes. Nevertheless, our GA based WFS strategy is appropriate for nonlinear systems as shown below.

**SRS and FWM in fibers**

SRS cascade generation is an important nonlinear process that builds up throughout the fiber from spontaneous Raman scattering. Phase matching for SRS is satisfied throughout the fiber since the medium is actively participating in the interaction in the sense that the process depends on lattice-vibrations of the fiber. The Raman gain, $g_R$, for fused silica is maximal at 13.2 THz (440 cm$^{-1}$), and therefore, the first Stokes line, at 440 cm$^{-1}$, builds up most rapidly once the power reaches the SRS threshold, and the energy is transferred from the pump to the Stokes wave. For sufficiently large input laser pulse power, before all the energy is transferred, the Stokes wave itself serves as a pump to generate a second order Stokes wave. If its power becomes strong enough, this process can generate a cascade SRS of multiple Stokes bands with its order increasing with fiber-length[18]. SRS cascades were demonstrated first using single mode[36] and small-core[37] fibers, and later on using large core MMF[38–41] and highly customized fibers[42].

Considering single mode propagation, the number of Stokes bands depends primarily on the input power. In MMFs the spatial overlap integral of the pump and Stokes along the fiber determines the efficiency of the interaction for a given input power. Stokes waves can evolve into one of the low-order modes or a combination of these modes under suitable light launching conditions (as shown below) and the efficiency of the process for a particular mode depends on the coupling efficiency of the pump into the individual mode[38].

Four wave mixing (FWM) is another dominant phenomenon that interplays with SRS in MMFs. It is a $\chi^3$ parametric nonlinear process that involves the interaction of four optical waves. Two pump waves annihilate to produce Stokes and anti-Stokes (frequency up shifted) photons. Here the medium plays a catalytic role and optical momentum conservation is required before nonlinearities can build up. In single mode fibers, there are several techniques for achieving phase matching[18]. However, the presence of multiple propagating modes in MMFs, each of them having different dispersive properties and corresponding momenta, results in expanded phase-matching combinations for the generation of FWM signals[43].

The WFS control of the interplay between SRS and FWM is the major phenomenon that we explore in this work. In long fibers, SRS practically dominates the interaction because it is difficult to maintain phase matching over long fiber lengths. In shorter fibers, the phase matching condition in MMF can be satisfied for several combinations of the fiber modes. As such, the multimode nature of these fibers presents numerous additional opportunities for exploiting modal-phase-matching to enhance nonlinear interactions that remain unexplored to date.

In order to illustrate the richness of phenomena, in our experiments we launch 532nm ns pulses into a GRIN MMF. Remarkably the SRS-FWM cascade generated extends from 470nm up to 1700nm (the limit of our detection) and possibly beyond; with all the peaks undergoing mode cleaning (see near-field images in Fig.1(b)), e.g they contain low order modes rather than the typical speckle fields of linear MMF. The generated cascade is stable with fiber movement and the stability increases with fiber length. This simple MMF system produces a highly nonlinear, tunable, multiple frequency single-mode source, that can be used for various application. In what follows we investigate the influence of the input wavefront on the generated nonlinear phenomena.

## WAVE-FRONT SHAPING CONTROL OF NONLINEARITIES IN FIBERS

Here we describe experiments for WFS control of nonlinear propagation in fibers. The optical setup is depicted in Fig. 1(a) and described in detail in methods. The key components are a nanosecond laser directed to a SLM, which modulates the light coupled into a MMF, and a spectrometer that provides feedback to the computer controlling the SLM.

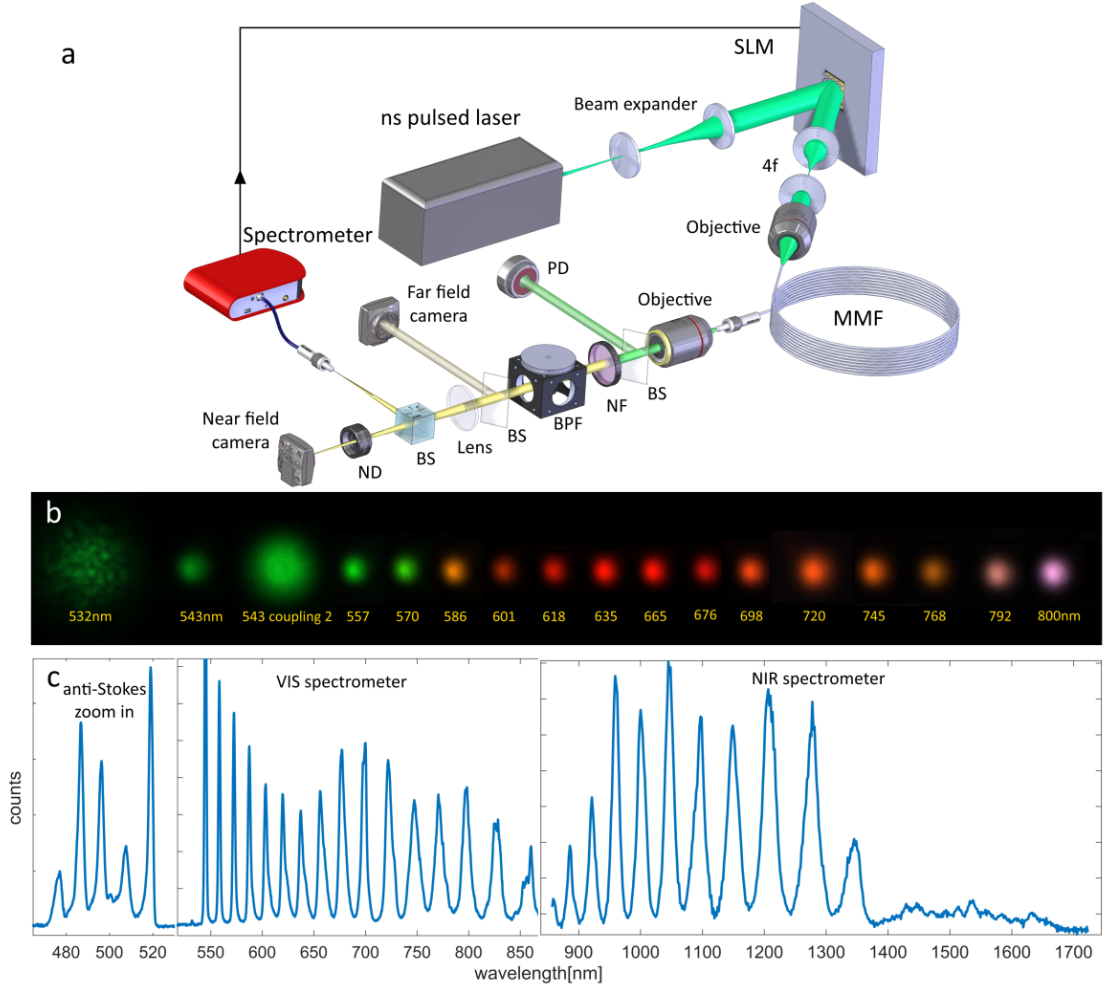

*Figure 1 : **System for wavefront shaping in nonlinear multimode fibers**. a- Optical setup (see methods for more details). PD - photodiode, NF - notch filter, BS - beam splitter, BPF - band pass filter, ND - neutral density filter. b- Near-field imaging of SRS cascade in a MMF. c- Spectrum of the SRS-FWM cascade in 1Km GRIN MMF. Anti-Stokes peaks are recorded using an additional short-pass filter. The filtering and integration time in the three regions of the spectrum were varied for representation. Pulse energy ~50μJ. Repetition rate 20KHz.*

For WFS, we divide the SLM into independent macro-pixels with phases varying between 0 and 2π. A genetic algorithm (GA) optimizes the values of each macro-pixel based on a merit function[44–46], as described in detail in methods. Accordingly, for each experiment the key merit function characterizes a specific spectral feature, which is recorded at the output tip of the fiber and fed back to the computer. The GA process starts with a set of random phase patterns and iteratively converges to an optimized phase mask that enhances the selected spectral feature.

In the first experiment, we investigate the enhancement of FWM interaction in short fibers. Using WFS, we optimize the intensity (maximum of the peak count value) of the first FWM anti-Stokes line at 517nm. We analyze the anti-Stokes side of the spectrum that contains only FWM peaks without the SRS peaks that dominate the Stokes side. The optimized SLM phase shows significant (six-fold) enhancement in the peak intensity compared to a flat phase at optimal mechanical focus alignment, a reference case in which the SLM serves as a mirror and the manual coupling maximizes the peak (Fig. 2). In the comparison, the input energy is kept constant and the flat phase spectra is measured before and after optimization to validate mechanical and thermal stability. The WFS optimization rises up sharply, at a certain threshold, reflecting the nonlinear nature of the feedback. Interestingly, the FWM anti-Stokes peak propagates as a LP$_{21}$ mode (Fig.2 (c)), completely different than the pump (multimode) and Stokes waves (mostly LP$_{01}$). Here the WFS optimization maximizes the FWM by launching a phased matched combination of pump modes.

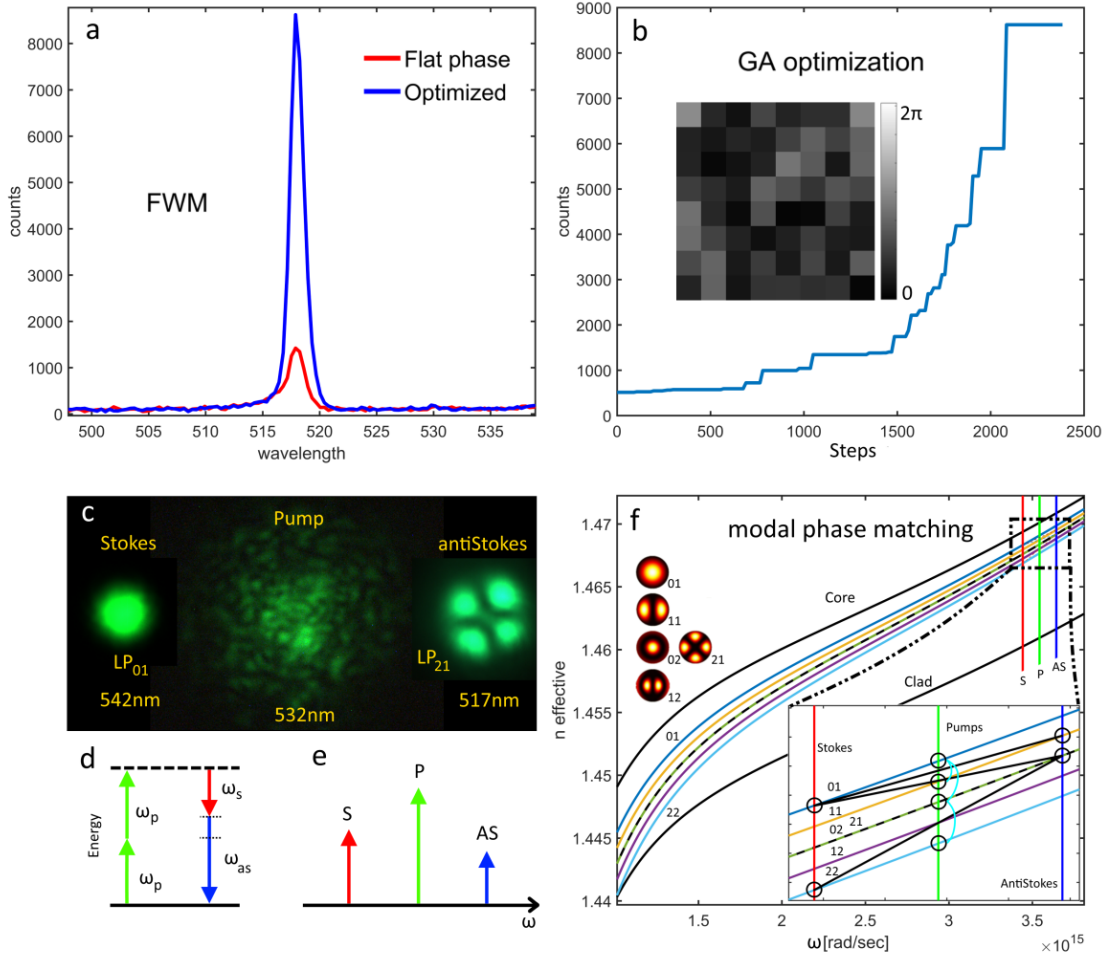

*Figure 2: **Wavefront-shaping of four-wave-mixing**. a – enhancement of FWM peak at 517nm. The 2.65m GRIN fiber was stretched in an aluminum v-grove rail and thermally stabilized in water-based gel. b – Optimization process. Maximal values at each step of the GA algorithm are depicted. The Insert shows the optimal phase-pattern. A 0.14NA objective was used with a pulse energy of 32µJ. c – Near-field image of the Stokes, pump, and anti-Stokes peaks at the fiber output. d,e – FWM energy and spectrum schemes. f - Simulation of intermodal-phase-matching. The material refractive-index is plotted (black) including material-dispersion[47]. Several modes in the GRIN fiber are shown in colors. The bottom-right inset is zoomed-in on the experiment regime. The red, green and blue spectral-lines denote Stokes(S), Pumps (P), and anti-Stokes (AS) wavelengths, intersecting with the calculated modes, and indicating possible propagating waves. The black lines indicate allowed phase-matched combinations. For each, we marked the two corresponding pumps with black circles and connecting light-blue curves. For each phase-matched process, the pumps average falls on the crossing of the black line and pumps spectral line.*

Intra-mode phase matching was first observed by Stolen et al[48,49] and recently regained new interest[14,43,50,51]. Our simulation of the material refractive-index of modes in GRIN fibers describes the mechanism of intermodal-phase-matching (Fig.2 (f)). Accordingly, the observed FWM peaks could be created by pump waves that satisfy the phase matching condition,

$$\Delta\beta = \beta_{01}^{s} + \beta_{21}^{as} - \beta_{lm}^{p} - \beta_{l'm'}^{p'} = 0 \qquad (1)$$

where $\beta_g^{wave} = n_{eff} k_0$ is the propagation constant of the modal-group-number, $g = |l| + 2m + 1$, and its k-vector. The index *wave* indicates either pump (*p*), Stokes (*s*) or Anti-Stokes (*as*). The optimized 532nm pump wave is highly multimode as indicated by its speckle pattern at the output and the WFS optimization maximizes the launching of phase-matched pump modes at the input. Note that a single mode pump at $LP_{01}$ cannot produce a $LP_{21}$ anti-Stokes with $LP_{01}$ Stokes mode through the phase-matched process. Such a combination also violates angular momentum conservation. Therefore, it is evident that the pump comprises higher order modes and, upon optimization, the SLM launches efficiently a combination of phased-matched modes into the fiber, systematically surpassing what is possible with manual coupling. The mechanism for phase-matching could include additional nonlinear effects, generating additional momenta along the GRIN fiber [17,52,53]. The complexity of these nonlinear interactions highlights the advantages of WFS optimization that accounts for all the dynamically-rich effects for a desired response.

Next, we investigate the Stokes side of the spectrum with the goal of enhancing the SRS cascade in a 100m GRIN fiber. The GA optimization merit function was set for the enhancement of a selected spectral region of interest. Fig.3 depicts the SLM control over of the cascade as we selectively tune the nonlinear interaction in the MMF.

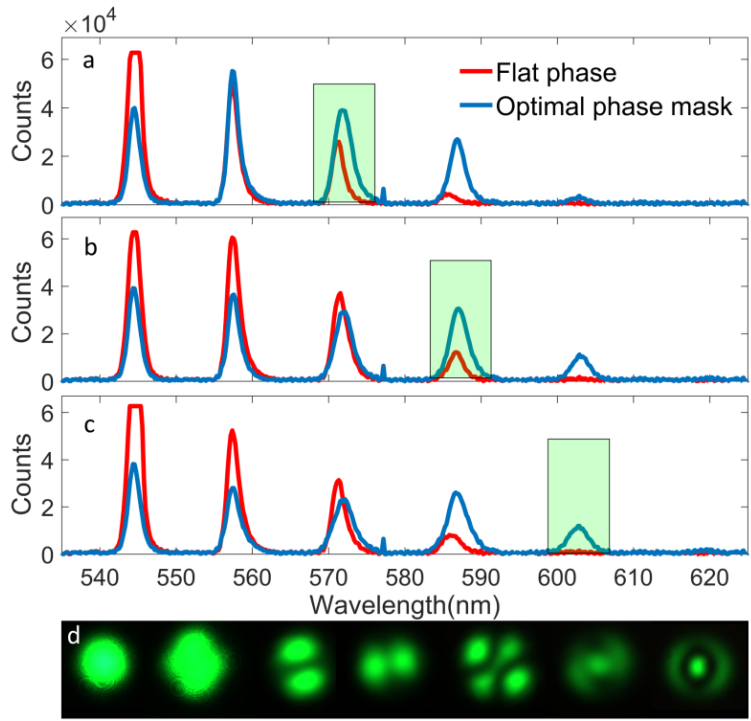

Figure 3: **Wavefront shaping of SRS peaks**. (a) 3rd SRS peak optimization results. (b) 4th SRS peak optimization. (c) 5th SRS peak optimization. The green square marks the spectral region of interest for each optimization. In each plot, a flat phase (red) on the SLM is compared to the optimal phase mask (blue). The input energy was kept constant for comparison. A 0.14NA objective was used with a pulse energy of 18μJ(d) Near-field images of SRS 542nm peak modes in a 20m GRIN fiber, each with different manual coupling.

The SRS cascade can be generated in several low order modes, as depicted in Fig.3 (d). However the most efficient cascade is generated once the fundamental mode is excited. In this case, the mode-cleaned pump overlaps spatially with the generated Stokes wave and the cascade keeps generating clean fundamental modes of higher wavelengths. The SLM optimizes the input superposition of modes for fundamental mode excitation, compensates for aberrations in the optical system, and enables dynamic feedback monitoring on the SRS cascade.

In terms of the modal control, the optimization of the SRS cascade efficiency is a rather simple example because it does not include complex modal excitation. On the other hand, the SRS interplay with FWM illustrates more complex intermodal phase matching. Such is the case in the next experiment where we generate a SRS cascade in a 1km long GRIN MMF and demonstrate spectral shifting of each peak of the cascade. The spectral shifts occur as the input excitation of the fiber is continually tuned from the optimized fundamental mode (longer SRS wavelengths) to a mixed mode excitation (SRS wavelengths downshift). At mixed modal excitation, the interplay of FWM becomes dominant and mediates the SRS cascade. For this experiment, we defined the WFS figure of merit function as the weighted average wavelength location in a selected spectral ROI as follows:

$$\lambda_{merit} = \lambda_{ref} \pm \frac{\int_{\lambda_1}^{\lambda_2} \lambda_n * I_{SRS_n}}{\int_{\lambda_1}^{\lambda_2} \lambda_n}, \quad (2)$$

where $\lambda_1$ and $\lambda_2$ define the spectral ROI, $I_{SRS}$ is the spectrum intensity and $\pm$ defines the wavelength shift direction with respect to a reference, $\lambda_{ref}$. The SLM provides continuous control over the spectrum to produce the desired output by controlling the mixture of modes at the input of the MMF.

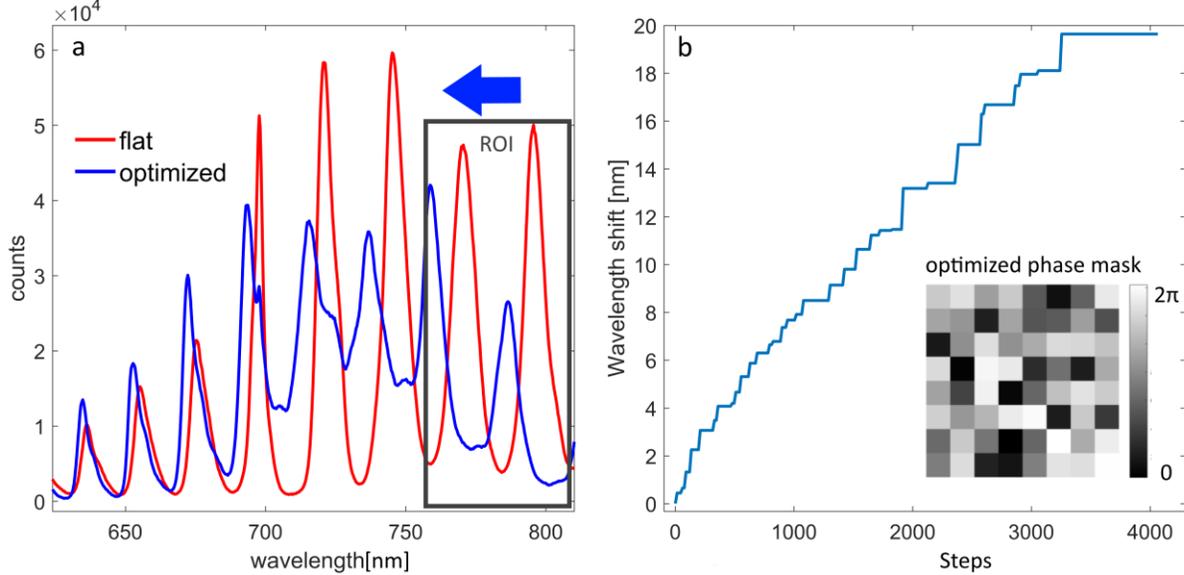

*Figure 4 **Wavefront shaping of spectral shifts**. a – Optimization of shifts towards lower wavelength and mixed mode excitation. The GA optimization process generated shifts of up to 20nm. The spectrum is continually shifted with the excitation of higher order modes that support FWM processes along the SRS cascade. The gray rectangle shows the selected spectral region of interest. b- The plot shows the GA performance and the bottom-right insert depicts the optimized phase mask. A 0.14NA objective was used with laser pulse energy of 50µJ.*

The spectral shifts of the cascade are obtained by mixed-mode excitation[38,54]. Here, the GA provides an optimized collection of modes on the SLM that generates a selective FWM interaction to pull the average wavelength down. Similar spectral-shifts can be achieved by manually adjusting the input coupling of the fiber[38,54]. However, the SLM provides a systematic and controlled feedback methodology that allows precise modal excitation for the desired results.

A tunable source based on nonlinear WFS control could be beneficial for various laser applications. However, for light-wave communications, nonlinearities limit the information capacity of fiber networks[55,56]. Suppression of nonlinearity is therefore highly desirable for the constantly growing bandwidth demand. In Fig.5 we demonstrate suppression of the SRS cascade in a 1Km GRIN fiber. The suppression feedback figure-of-merit, $F_{merit}$, comprises two components: the total energy of the SRS cascade, $I_{SRS}$, and the total transmission in the fiber as follows

$$F_{merit} = \frac{1}{2}\left(\frac{\int_{\lambda_1}^{\lambda_2} I_{SRS\ (ref)}}{\int_{\lambda_1}^{\lambda_2} I_{SRS}} + \frac{I_{out}}{I_{out(ref)}}\right), \quad (3)$$

where $\lambda_1$ and $\lambda_2$ are the spectral limits of the ROI. $I_{SRS}$ appears inverted and normalized, while $I_{out}$ is normalized. The fiber transmission, $I_{out}$, is measured at the output before spectral filtering and its place in the figure-of-merit assures that the suppression of SRS is the result of high-mode excitation and not simply decoupling of light by diffraction on the SLM. For simplicity, we weight the two optimization components equally but it is possible to choose a different weighting function.

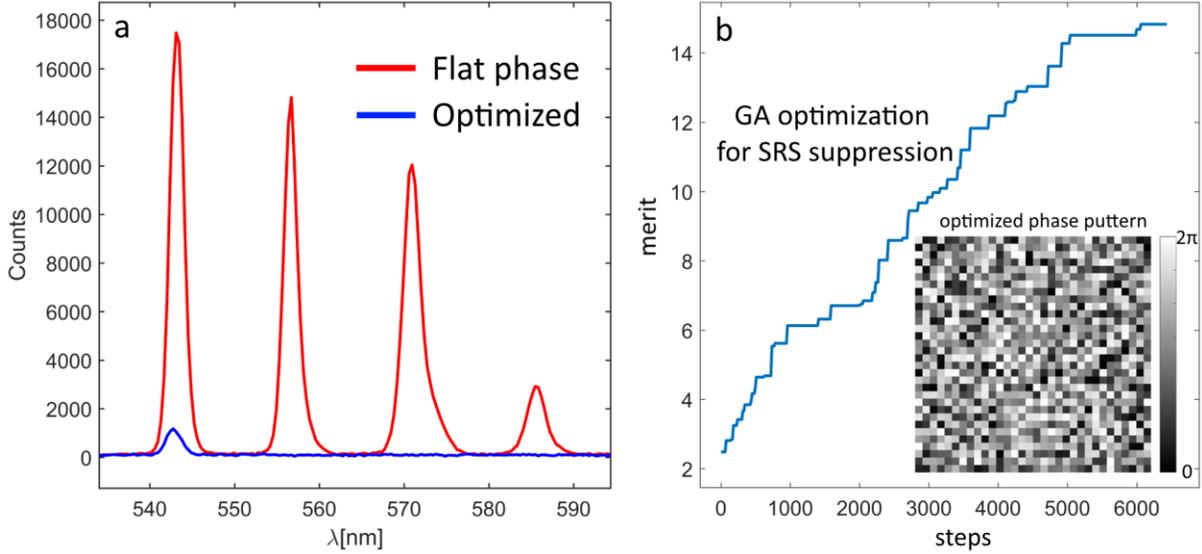

*Figure 5 : **SRS cascade suppression through high mode excitation**. a –comparison of a normalized flat-phase (with 66% reduced energy) and the optimized-phase spectra with matched total transmission. b- Performance of the GA suppression optimization. Inset- Phase mask for joint optimization of suppression through higher order mode excitation and transmission. A 0.25NA objective was used, the pulse energy 22.5µJ.*

In addition, we measure the total transmission before and after optimization to normalize any diffraction effects in the suppression experiment. After optimization, the projected phase pattern decreased the total transmission by 10%, compared to an averaged random-phase (initial mask in the GA process) and by 66%, compared to the flat-phase-transmission. In the flat-phase comparison of Fig.5 (a), we reduced the laser energy by 66%, compared to the power used in the optimization, and plotted the flat-phase output with identical total transmission to the optimized-phase mask. Our optimization suppressed the SRS cascade by a factor of x52. This value represents the ratio of the integrated SRS cascade spectra for the normalized flat-phase and optimized-phase cases. Fig. 5(b) shows the WFS enhancement of the figure-of-merit with iteration number. The dramatic, x7, suppression of the SRS cascade from the initial random pattern shows that optimizing the higher order mode superposition is significantly more effective than a non-optimized speckle pattern, such as could be achieved with a simple NA-matched diffuser.

## DISCUSSION

The WFS control of various nonlinear processes presented here enables spectral shaping via the coupled spatial modal control. Generally, as the fibers get shorter, the effect of WFS becomes more noticeable. Fiber cut-back experiments show that <50m fibers support several SRS modes while in the longer fibers, mostly the fundamental clean mode appears, hinting that a mode competition occurs along the length of the fiber. Even with short fibers of <5m, the most efficient SRS cascades occur once the fundamental mode is excited efficiently. While similar effects could also be achieved without a SLM, using an optimized lens coupling into the fundamental mode, WFS provides a controlled way for enhancing the mode excitation. Furthermore, WFS enables control over FWM, spectral shifting, and nonlinearities suppression by coupling a tailored superposition of modes into the fiber. All of these phenomena and capabilities are attained beyond the capabilities of simple lens-coupling.

GRIN fibers have unique properties for generating interesting nonlinear interactions not always shared by step-index fibers. For instance we tested step-index fibers (50 microns, 10m) and could not attain SRS nor FWM with our maximal laser power.

The number of macro-pixels utilized in the SLM has to be carefully considered. Once phase patterns are displayed, some of the light is diffracted out of the fiber, reducing the coupled input power and decreasing the nonlinearity regardless of the particular modes excited. As the number of SLM macro-pixels increases, the diffraction spectrum broadens, further reducing the input power coupling. In order to limit this unwanted diffraction, the number of SLM macro-pixels was limited between 64–1024, and each phase pattern was slightly low-pass digitally filtered to smooth the phase edges. We took into account the diffraction decoupling effects and presented our results with strict criteria, namely a constant laser power for enhancement and a normalized coupling power into the fiber for suppression, as described above.

Thermal management plays an interesting role in the SRS cascade. For the WFS experiments, we were mostly concerned about the coupling of thermal effects with optimization of the efficiency of nonlinearities. To eliminate thermal effects, we aligned a short fiber in an aluminum v-shaped profile and immersed the fiber in water-based gel. This configuration allowed improved thermal management, and we compared the flat phase spectrum before and after optimization to ensure that the thermal management keeps the fiber at the same conditions over time and during WFS. Note also that liquid-crystal SLMs and other phase modulators are subject to optical damage in high-power applications. Proper precautions ensure safe operation with a high-power ns laser.

The application of nonlinear MMF requires long term stability of the system. Using an active device, such as an SLM allows degree of dynamic control that maintain operation over long periods of time compensating for mechanical and thermal drifts. Such techniques open up a new field of adaptive nonlinear optics.

In moving forward and generalizing nonlinear WFS it is interesting to explore how WFS controls systems with different types of nonlinearity. We preformed our experiment in a specific regime: wavelength of 532nm, normal dispersion in the fiber, and ns pulses where group-velocity-dispersion effects are of minor importance. We demonstrated several important applications of WFS control in nonlinear MMF including enhancement, shifting and suppression of SRS and FWM. We expect similar WFS methodologies to be even more significant in other systems, for instance as femtosecond pulses in the anomalous dispersion regime are considered.

## Conclusion

We presented WFS control and optimization of nonlinear interactions in MMF. By optimizing the input phase of the fiber-coupled wavefront, we tuned the energy of selected SRS and FWM peaks creating a configurable source with tailored performance. The adaptive in-line optimization represents an approach to solve the nonlinear inverse problem of finding a tailored superposition of modes at the input of the fiber for a desired spectral output. This work opens opportunities for characterizing and controlling rich spatiotemporal dynamics in MMF. Its potential applications include nonlinear frequency generation, high power MMF lasers, nonlinear endoscopy, and nonlinearity suppression in multimode-fibers.

## Methods

### Experimental setup

The optical setup used in the experiment is depicted in Fig. 1(a). It includes a laser source (Spectra Physics, Mosaic) with 532nm, ~7ns pulses, energy up to 150μJ and repetition rate of 20 KHz. Wave-front shaping is performed using a liquid crystal spatial light modulator (LC-SLM) (Meadowlark 512x512). Reflected light of the SLM is imaged by a 4f system onto the back aperture of a microscope objective (Olympus, 0.14 NA or Leica 0.25NA) that couples the light into the MMF. We used an off-the-shelf GRIN fiber (Corning, 62.5/125-μm) with changing lengths in the range of 2.65m – 1Km. For the SRS enhancement experiments (Fig. 3(a)-3(c)), we used a 100m, 62.5/125-μm fiber (Thorlabs GIF625). Step-index fiber (Thorlabs FG050LGA) was used as well. The fiber output was coupled into a customized microscope for near and far-field imaging of fiber modes with the possibility to switch between different optical configurations. The microscope includes output power monitoring, notch filter (Thorlabs NF533-17), tunable ND filter attenuation, and a series of narrow band pass filters. The anti-Stock bands were analyzed using a short-pass filter (Semrock BPS01-532-25) to avoid saturation of the detector due to the intense SRS. For parallel spectral detection, the light beam was split and coupled into a multimode-fiber to average the spectrum in space and from there coupled into a spectrometer (OceanOptics Flame VIS-NIR or NIR-512). The signal from the spectrometer is acquired and analyzed by a computer. For WFS, we divide the SLM into independent macro-pixels whose phase varies between 0 and 2π. A genetic algorithm (GA) optimizes the values of each macro-pixel based on a merit function tailored to the experiment.

**Genetic Algorithms**

The GA optimization for WFS[44,45] starts with a population set of random phase masks (30 in our case) and iteratively converges to an optimized pattern. At each step, a phase mask is displayed on the SLM and a merit signal is recorded, based on a specified spectral analysis at the fiber's output. The recorded values of the initial population are ranked based on the selected figure-of-merit and a new generation, containing new phase masks (15 off-springs in our case) is created. The breeding process combines two phase masks from the population, which are randomly chosen with a probability weighted by the ranking. At each step of the GA, a new phase mask is displayed on the SLM and the corresponding figure-of-merit is recorded. The ranking and breeding process repeats itself every cycle (15 measurements in our case), always keeping the 30 highest ranked phase-masks as the population for the next iteration. As a result, a phase mask is found that enhances the selected figure-of-merit and corresponding spectral feature. The optimization time is determined by the number of steps and the acquisition time. Typically, the spectrometer acquisition time is set to 10ms with additional x3 averaging for each step. The SLM refresh time is also in the order of 10ms leading to approximately 50ms per measurement. Accordingly, for ~10,000 measurements, the optimization times were in the order of 10 minutes using a non-optimized MATLAB software.

**Funding.** We acknowledge support from the National Science Foundation through award 1611513 and from the National Institute of Health award REY026436A.

**Acknowledgment.** We thankfully acknowledge fruitful discussions with Frank Wise, Logan Wright, and Ronald Ulbricht at the early stages of this work. We thankfully acknowledge Sakshi Singh for help with the mode simulations.